\documentclass[prb,aps,reprint,english,twocolumn,notitlepage]{revtex4-1}
\usepackage[T1]{fontenc}  \usepackage[latin1]{inputenc}
 \usepackage{babel} \usepackage{txfonts}
 \usepackage{epsfig,epsf,psfrag} 
\usepackage{graphicx} 
\usepackage{subfigure}
\usepackage{pslatex}
\usepackage{epic,eepic} 
\usepackage{color,pstcol}
\usepackage{pstricks}
\usepackage{fancyhdr}  \parindent0em  

 \vspace{-10.0cm}  
\begin{document} \title{Strained graphene based highly efficient quantum heat engine operating at maximum power} 
  \author{Arjun Mani}
 \author{Colin Benjamin} \email{colin.nano@gmail.com}\affiliation{School of Physical Sciences, National Institute of Science Education \& Research, HBNI, Jatni-752050, India}
\begin{abstract}
A strained graphene monolayer is shown to operate as a highly efficient quantum heat engine delivering maximum power. The efficiency and power of the proposed device exceeds that of recent proposals. The reason for these excellent characteristics is that strain enables complete valley separation in transmittance through the device, implying that increasing strain leads to very high Seebeck coefficient as well as lower conductance. In addition, since time-reversal symmetry is unbroken in our system, the proposed strained graphene quantum heat engine can also act as a high performance refrigerator. 
 \end{abstract}
\maketitle
\section{Introduction}
Quantum heat engines(QHE) have twin purposes to act as highly efficient heat engine enabled by quantum principles and also to act as a conduit for excess heat \cite{selman}, therein lies their appeal. In this work we probe the thermo-electric properties of an open strained graphene system and show its action as a very efficient quantum heat engine(QHE) \cite{prosen}. The heat engine of ours is a steady-state heat engine. Steady-state devices convert heat to work without any macroscopic moving parts, through steady-state (no time dependent parameter involved) flows of microscopic particles such as phonons, electrons and photons\cite{benenti,ron}. Heat engines can not only be of steady state type but also be classified as cyclic heat engines, such as Carnot and Otto engines. These cyclic heat engines are non-autonomous, since they require an external control system, while the steady state heat engines are autonomous. In cyclic heat engines all parameters revert to their initial position in one period, drawing thereby a cycle in parameter space, through a reversible transformation. Reversible operations imply extremely long duration of the working cycle and as a result when the engine efficiency reaches Carnot efficiency the output power is zero. In a steady-state device one can get finite output power with an efficiency below the upper bound of Carnot efficiency. One of such cyclic/closed heat engines, working in graphene, is mentioned in Ref.~\onlinecite{munoz}. A closed system differs from an open system in that no transport of heat or charge current is involved in the operation of such a closed heat engine, so there is no dissipation within the system, and no power generation too \cite{benenti}.

Recent works show that graphene is a good thermoelectric material with a moderate Seebeck coefficient\cite{dragoman, gahari}. Though, due to its high thermal conductance, it possesses a very small thermoelectric figure of merit ZT around 0.1-0.001, much smaller than the most efficient 3D thermoelectric material $Bi_2Te_3$. In some  recent works based on graphene, ZT's close to 3 have been obtained in presence of disorder\cite{anno, tran} or nanopores\cite{sadeghi} or isotopes\cite{tran} or by nanopatterning the graphene surfaces\cite{yun} at room temperature. The maximum output power at this ZT value is still smaller than that of Ref.~\onlinecite{sothmann}, which is a quantum Hall heat engine working at much lower temperatures. 

In this work, we consider the possibility of strained graphene as an efficient thermo-electric material in the ballistic transport regime. The presence of uniaxial strain in graphene, introduces quantum confinement by suppressing the transmission at particular incident angles, which, in turn, increases the Seebeck coefficient to large values even at lower temperatures, though, it reduces the electrical and thermal conductances. In this work we see that the dimensionless thermoelectric figure of merit ZT can increase to a value as large as 3 even at temperatures around 30 K, as well as the maximum output power can be increased, which is now comparable to that of Ref.~\onlinecite{sothmann} and demonstrates the potential of graphene as an efficient heat engine. In Ref.~\onlinecite{dolphas}, they have generated a large Seebeck coefficient, in a similar device setup to us, but in a different context, to use it as a thermal sensor rather than a quantum heat engine. The rest of the paper is organized as follows- in section II we deal with the theoretical framework of our quantum heat engine and derive the power, efficiency and thermoelectric figure of merit while in section III we describe our model and calculate the scattering coefficients needed to determine the efficiency and power of our strained graphene based QHE. Section IV  is devoted to the results of our work wherein we also discuss the reasons for obtaining the large values of efficiency and power of our QHE. A notable aspect of our work is that our strained graphene model can also be effectively used as a quantum refrigerator in  a different parameter space we  discuss the coefficient of performance of the quantum refrigerator in section V. We end our paper with a Conclusion where we put our work in perspective and compare our results with those of some other proposals in  Table 1.
\\
\begin{figure}
\includegraphics[width=0.5\textwidth]{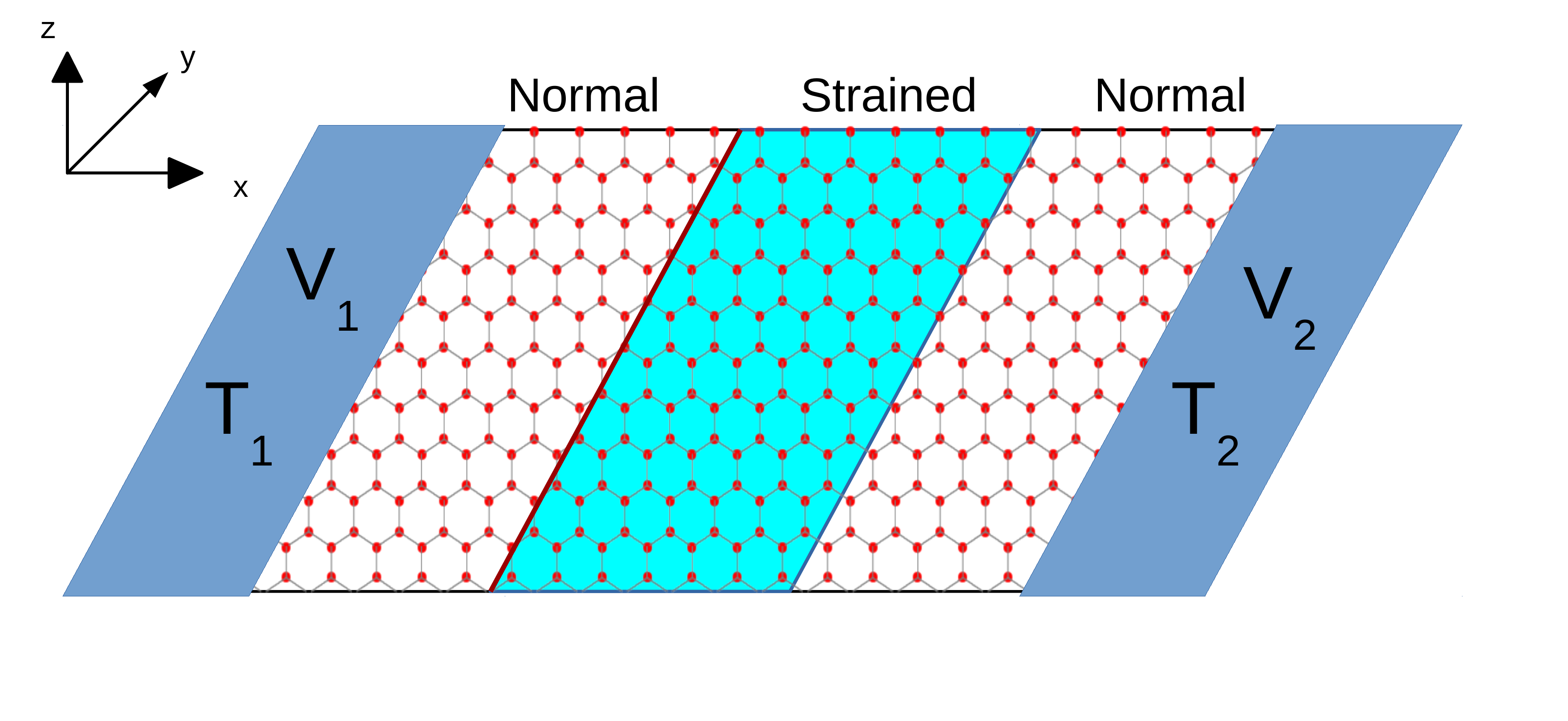}\\
{\vspace{-0.38cm}}
\includegraphics[width=0.5\textwidth]{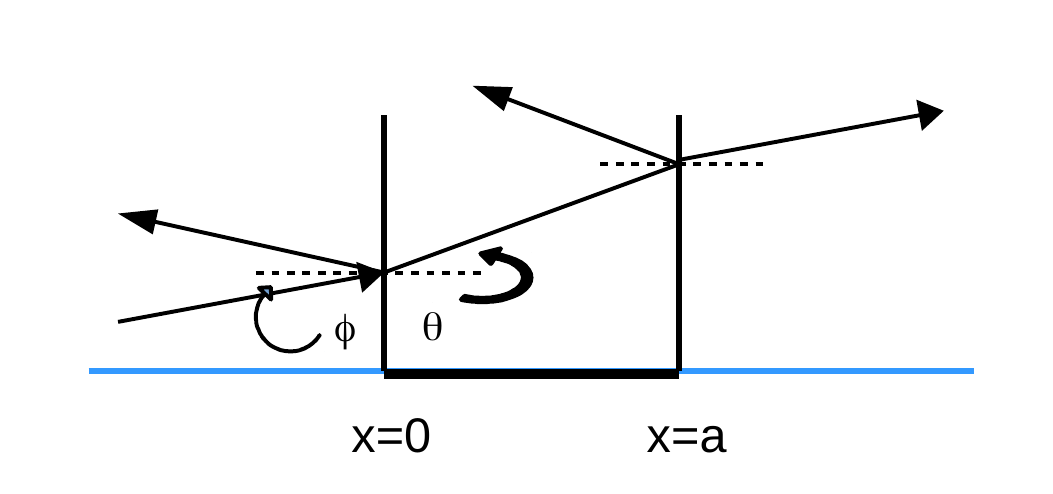}
\caption{Top: The graphene layer with a strained region between $x=0$ and $x=L$. Electric and heat currents are generated due to the applied potential bias and finite temperature difference between probe 1 and 2. Bottom: A incident electron at the interface between region 1 and 2 is transmitted or reflected by the strained region with a finite probability. Here $\phi$ is the incident angle and $\theta$ is the refracted angle in the strained region.}
\end{figure}
\section{Theory}
Our aim in this work is to design an extremely efficient QHE which operates at full power using a strained graphene system. To do this we calculate the thermoelectric properties of our system in the linear transport regime, wherein electric and heat currents are linearly proportional to the applied biases be it electric or thermal. In a thermoelectric system temperature difference $\Delta T$ and applied electric bias $\mathcal{E}$ across it work in tandem to operationalize it. The linear dependencies can be expressed as follows \cite{ramshetti, benetti, sothmann}-
\begin{equation}\label{current}
\left(\begin{array}{c} j \\ j^{q}\end{array}\right)=\left(\begin{array}{cc} L^{11} & L^{12}\\ L^{21} & L^{22} \end{array}\right) \left(\begin{array}{c} \mathcal{E}\\ \Delta T \end{array}\right)
\end{equation}
where $j$ and $j^q$ are the electric and heat currents respectively, $L_{ij}$ with i,j $\in$ 1,2 represents Onsager coefficients for a two terminal thermo-electric system. 
The Seebeck coefficient is defined as the electric response due to the finite temperature difference $\Delta T$ across the system. On the other hand, the Peltier coefficient $P$ is defined as the heat current generated due to the applied bias voltage $\mathcal{E}$ across the system. They are expressed as follows-
\begin{equation}\label{Seeback}
S=-\frac{L^{12}}{L^{11}}, \quad\text{and}\qquad P=\frac{L^{21}}{L^{11}}
\end{equation}
The Onsager co-efficient matrix in Eq.~[\ref{current}], which relates the electric and heat currents to the temperature differences and applied electric bias, can be rewritten as follows \cite{ramshetti, mazzamuto}-
\begin{equation}\label{onsager}
\left(\begin{array}{cc} L^{11} & L^{12}\\ L^{21} & L^{22} \end{array}\right) =\left(\begin{array}{cc} \mathcal{L}^{0} & \mathcal{L}^{1}/eT\\ \mathcal{L}^{1}/e & \mathcal{L}^{2}/e^2T \end{array}\right)
\end{equation}
wherein,
\begin{eqnarray}\label{con}
\mathcal{L}^{\alpha}=G_0\int_{-\pi/2}^{\pi/2}\!\!\!d\phi \cos\phi \int_{-\infty}^{\infty}\!\!\!d\epsilon(-\frac{\partial f}{\partial \epsilon})\frac{|\epsilon|}{\hbar v_f}(\epsilon-\mu)^\alpha T(\epsilon,\phi)
\end{eqnarray}
here $G_0=(e^2/\hbar)(W/\pi^2)$, $\mathcal{L}^0=G$ is conductance of system with sample width $W$ \cite{dolphas}, $\phi$ is the angle at which the electron is incident, $\epsilon$ is the energy of the electron, $f$ is the Fermi-Dirac distribution, $\mu$ is the Fermi energy and $T(\epsilon, \phi)$ is the transmission probability for electrons through strained graphene. To calculate the Onsager coefficients $L^{ij}$ in Eq.~(\ref{current}),  we need the transmission probability $T(\epsilon, \phi)$. After calculating the Onsager coefficients $L^{ij}$ in Eq.~(\ref{current}), maximal efficiency and power can be determined, as follows.
The output power \cite{benetti}, defined as -
\begin{eqnarray}\label{power}
\mathcal{P}=j\mathcal{E}=(L^{11}\mathcal{E}+L^{12}\Delta T)\mathcal{E} 
\end{eqnarray}
is maximized by $\frac{d\mathcal{P}}{d\mathcal{E}}=0$, at $\mathcal{E}=-\frac{L^{12}}{2L^{11}}\Delta T$, which gives maximum power as-
\begin{eqnarray}\label{pmax}
P_{max}= \frac{1}{4}\frac{(L^{12})^2}{L^{11}}(\Delta T)^2=\frac{1}{4}S^2G(\Delta T)^2
\end{eqnarray}
The efficiency at maximum power is defined as the ratio of maximum power to the heat current transported and is derived to be- 
\begin{eqnarray}\label{etapmax}
\eta(\mathcal{P}_{max})= \frac{\mathcal{P}_{max}}{j^q}=\frac{\eta_c}{2}\frac{T{L^{12}}^2}{2L^{11}L^{22}-L^{12}L^{21}}=\frac{\eta_c}{2}\frac{GS^2T/\kappa}{2+GS^2T/\kappa}\nonumber\\
\end{eqnarray}
at $\mathcal{E}=-\frac{L^{12}}{2L^{11}}\Delta T=\frac{S}{2}\Delta T$, which is the condition for maximum power, and $\kappa$ is the thermal conductance, defined as-
\begin{eqnarray}\label{kappa}
\kappa=\frac{L^{11}L^{22}-L^{12}L^{21}}{L^{11}}
\end{eqnarray}
Similarly, efficiency $\eta$ becomes\cite{benetti}-
 \begin{eqnarray}\label{eta}
\eta=\frac{\mathcal{P}}{j^q}=\frac{(L^{11}\mathcal{E}+L^{12}\Delta T)\mathcal{E} }{(L^{21}\mathcal{E}+L^{22}\Delta T)}=\frac{-(\mathcal{E}-S\Delta T)\mathcal{E}}{(TS\mathcal{E}-(\frac{\kappa}{G}+TS^2)\Delta T)}.
\end{eqnarray}
To calculate maximal efficiency we need to find the relation between $\mathcal{E}$ and $\Delta T$, substituting $\frac{d\eta}{d\mathcal{E}}=0$ in Eq.~(\ref{eta}), with the condition $j^q>0$, gives-
\begin{eqnarray}
\mathcal{E}&=&\frac{L^{22}}{L^{21}}(-1+\sqrt{\frac{L^{11}L^{22}-L^{12}L^{21}}{L^{11}L^{22}}})\Delta T\\
\text{and,}\quad\eta_{max}&=&\eta_c\frac{\sqrt{ZT+1}-1}{\sqrt{ZT+1}+1},
\end{eqnarray}
wherein $\eta_c$ is the Carnot efficiency defined by $\frac{\Delta T}{T}$ and $ZT$ is the figure of merit, a dimensionless quantity, defined as-
\begin{eqnarray}\label{ZT}
ZT=\frac{G S^2 T}{\kappa}
\end{eqnarray}

\section{Model}
\subsection{Hamiltonian}
Graphene is a 2D Carbon allotrope with honeycomb lattice structure which consists of two triangular sublattices A and B. To design our system we apply an uniaxial mechanical strain \cite{castro} to the monolayer graphene sheet lying in the $xy$ plane between $x=0$ and $x=L$. A potential bias is applied at contact 1 with a finite temperature difference between the two contacts 1 and 2. The corresponding set up is shown in Fig. 1. 
In Fig.~1, a general two terminal thermodynamic model is shown to operate between two temperatures $T_1>T_2$ and a bias $\Delta V=V_1-V_2$. At steady state, a steady heat and electric current, $j^q$ and $j$ flow between these two reservoirs. If $j^q>0$ and output (as defined in Eq.(\ref{power})) $power>0$ then it is a QHE and if $j^q<0$ and output $power<0$ then it acts as a refrigerator.

In Landau gauge, the strain can be expressed as a pseudo magnetic vector potential $A=(0, \pm A_y)$, where `+' and `-' signs are for $K$ and $K'$ valley respectively \cite{castro}.  
This system is described by the Hamiltonian, which is given for $K$ and $K'$ valleys as-
\begin{equation}\label{ham}
\mathcal{H}_{K}=\hbar v_f\sigma(k-s) \quad \mathcal{H}_{K'}=-\hbar v_f\sigma^*(k+s)
\end{equation}
Here $s=\frac{A_y}{\hbar v_f}[\Theta(x)-\Theta(x-L)]$ is the strain, $\sigma=(\sigma_x, \sigma_y)$ are the Pauli matrices operating on the sublattices A and B with $\sigma^*$ being the complex conjugate, $k(=\{k_x, k_y\})$ is the 2D wave vector, $\Theta$ being the step function and $v_f$ the Fermi velocity. Solving the Hamiltonian in Eq.~(\ref{ham}) we can write the wave equation for $K$ valley as-
\begin{eqnarray}\label{wav}
\hbar v_f(-i\partial x-\partial y-i s)\psi_B=E\psi_A\nonumber\\
\hbar v_f(-i\partial x+\partial y+i s)\psi_A=E\psi_B
\end{eqnarray}
In the next subsection we will solve the (\ref{wav}) to calculate the transmission $T(\epsilon, \phi)$ for ballistic transport in monolayer graphene with uniaxial strain.
\subsection{Wave function and Boundary conditions}
Let us consider an electron with energy $\epsilon$ incident on the interface between region 1 and 2 with angle $\phi$, which can reflect or transmit depending on it's energy and angle of incidence. Here, we have three well defined regions-normal graphene $x<0$, strained graphene between $x=0$ and $x=L$ and again normal graphene for $x>L$. The wave functions for the three regions for A and B sublattices in $K$ valley are given below.\\ For, $x<0$-
\begin{eqnarray}\label{a}
\left[\begin{array}{c}\psi_A^1(x,y)\\\psi_B^1(x,y)\end{array}\right]=\left[\begin{array}{c} (e^{ik_xx}+r e^{-ik_xx}) \\(e^{ik_xx+i\phi}-r e^{-ik_xx-i\phi})\end{array}\right]e^{ik_yy}
\end{eqnarray}
\\in region $0<x<L$-
\begin{eqnarray}\label{b}
\left[\begin{array}{c}\psi_A^2(x,y)\\\psi_B^2(x,y)\end{array}\right]=\left[\begin{array}{c} (a e^{iq_xx}+b e^{-iq_xx}) \\ (a e^{iq_xx+i\theta}-b e^{-iq_xx-i\theta}) \end{array}\right]e^{ik_yy}
\end{eqnarray}
and for $x>L$-
\begin{eqnarray}\label{c}
\left[\begin{array}{c} \psi_A^3(x,y)\\\psi_B^3(x,y)\end{array}\right]=\left[\begin{array}{c} t e^{ik_xx} \\t e^{ik_xx+i\phi} \end{array}\right]e^{ik_yy}
\end{eqnarray}
\\
where $q_x=\sqrt{(\epsilon/\hbar v_f)^2-(k_y-s)^2}$ is the $x$ component of momentum wave vector inside the strained region. In the normal regions $q_x$ is replaced with $k_x$ and $k_x^2+k_y^2=(\epsilon/\hbar v_f)^2$ wherein $k_x=(\epsilon/\hbar v_f) \cos \phi$ and $k_y=(\epsilon/\hbar v_f) \sin \phi$. In the strained region $q_x=(\epsilon/\hbar b_f) \cos \theta$ and $k_y-s=(\epsilon/\hbar v_f) \sin \theta$, $\theta$ being the refraction angle in the strained region as shown in Fig. 1(bottom) and also satisfies $\tan \theta=(k_y-s)/q_x$. To solve Eq.~(\ref{wav}) for the wave functions in Eqs.~(\ref{a}-\ref{c}) we impose following boundary conditions-\\
at $x=0$-
\begin{eqnarray} \label{ba}
\psi_B^2(x=0)=\psi_B^1(x=0), \quad
\psi_A^2(x=0)=\psi_A^1(x=0)
\end{eqnarray}
and at $x=L$- 
\begin{eqnarray} \label{bb}
\psi_A^2(x=L)=\psi_A^3(x=L), \quad \psi_B^2(x=L)=\psi_B^3(x=L).
\end{eqnarray}
Solving Eqs.~(\ref{ba}-\ref{bb}) we get the transmission probability for $K$ valley as- 
\begin{eqnarray} \label{t}
T(\epsilon,\phi)=\frac{1}{\cos^2[q_xL]+\sin^2[q_xL](\frac{1-\sin[\theta]\sin[\phi]}{\cos[\theta]\cos[\phi]})^2}
\end{eqnarray}
Finally from the Hamiltonian for $K'$ valley as in Eq.~(\ref{ham}) and imposing boundary conditions similar to that for $K$ valley and then replacing $\phi\rightarrow-\phi$, $s\rightarrow-s$ we get the transmission probability for $K'$ valley. The total conduction then is sum of both K and $K'$ valley conductances. It so turns out that although transmission $T(\epsilon,\phi)$ differs in $K$ and $K'$ valley, when integrated over $'\phi'$ this differences disappear. Thus total conductance $G$ is the twice that of $K$ valley conductance.

 \begin{figure}
\centering {\includegraphics[width=0.48\textwidth]{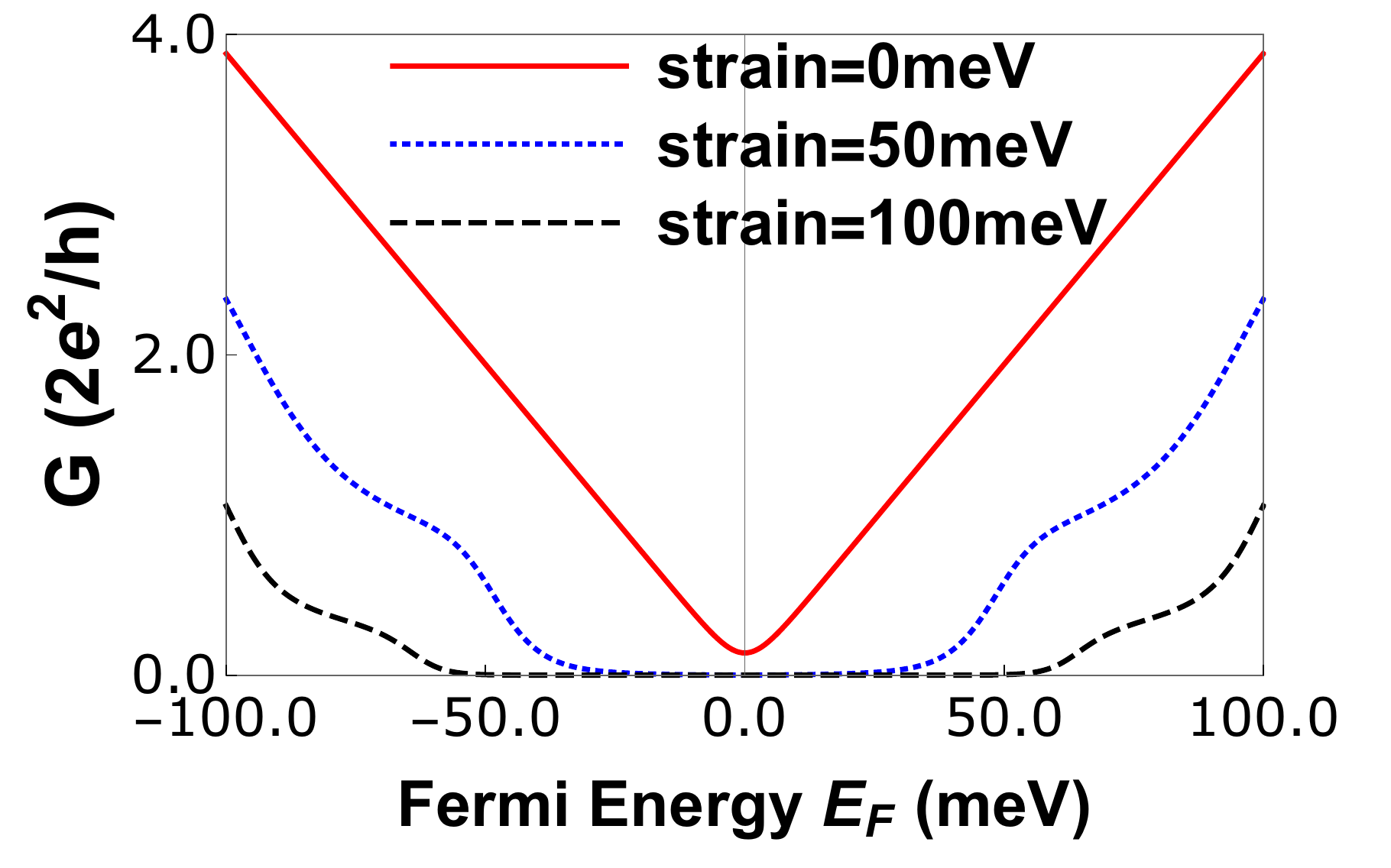}}
     \caption{Conductance (in units of "$2e^2/h$") at $30 K$ for various values of strain with $L=40 nm$ and width $W=20 nm$.}
\end{figure}
 \begin{figure}
  \centering {\includegraphics[width=0.48\textwidth]{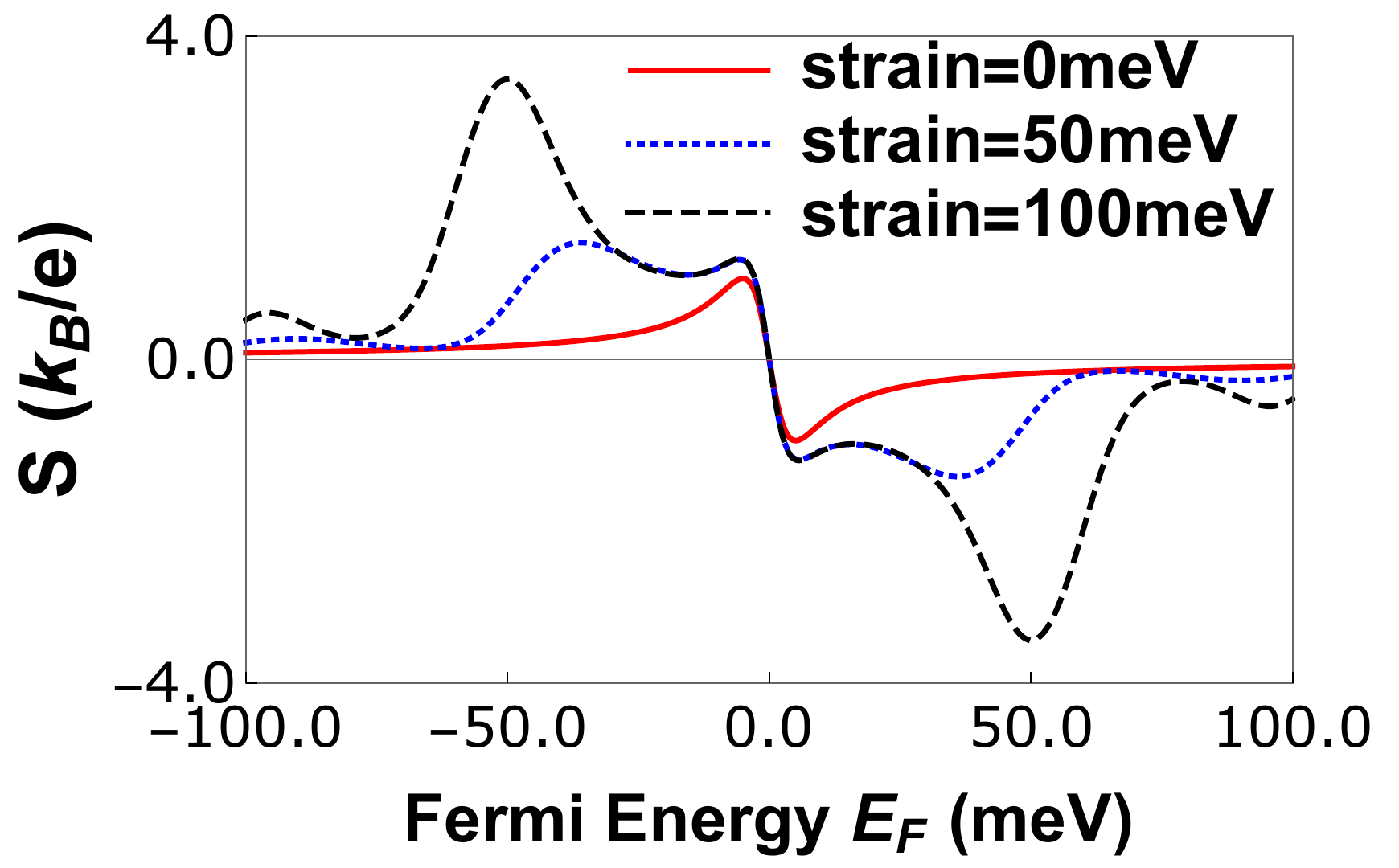}}
 \caption{Seebeck co-efficient S in units of ($k_b/e$) at $30K$ for different values of strain with $L=40 nm$ and width $W=20 nm$.}
\end{figure}

\section{Results and Discussion}
Our aim as defined in the introduction was to design an efficient QHE operating with full power using strained graphene. To do that we have to have high efficiency at maximum power. The generated power should be comparable to or better than other QHE's based on quantum Hall effect\cite{sothmann}, chaotic cavities \cite{buti} etc. To get maximum power the system should have a large Seebeck coefficient ($S$) with a large electrical conductance ($G$), as power is proportional to the $S^2G$, see Eq.~(\ref{pmax}).

Increasing strain reduces the electrical conductance, see Fig.~2, but increases the Seebeck coefficient, as in Fig.~3, which is also seen in Ref.~\onlinecite{dolphas}. As strain is increased the total transmission probability of electrons decreases, thus reducing the electrical conductance. From Fig~2, we see that increasing strain opens a gap in the conduction, though it is not a band gap, it is due to the shift of the Dirac cones by the strain in the Brillouin zone. A band gap opens for strain beyond 20 percent (540 meV) in pristine graphene \cite{perera}, so we will restrict ourselves only to a maximum of 15 percent strain (400 meV). A sign changed in Fig.~3, observed in the Seebeck co-efficient near the charge neutrality point (CNP), is due to switches between the carrier from hole to electron. The first peak, close to the CNP, is due to the imbalance of electron and hole contribution to the thermo-electric co-efficient $L^{12}$, presents even at zero strain, dies at a distance from the CNP. Though the origin of the second peak in the Seebeck co-efficient (blue line in Fig. 3) is the strain. As a result of applied strain transmission probability becomes a function of energy, and give rise to a large Seebeck co-efficient, which leads to a large power with a finite efficiency.
 \begin{figure}
  \centering{\includegraphics[width=0.48\textwidth]{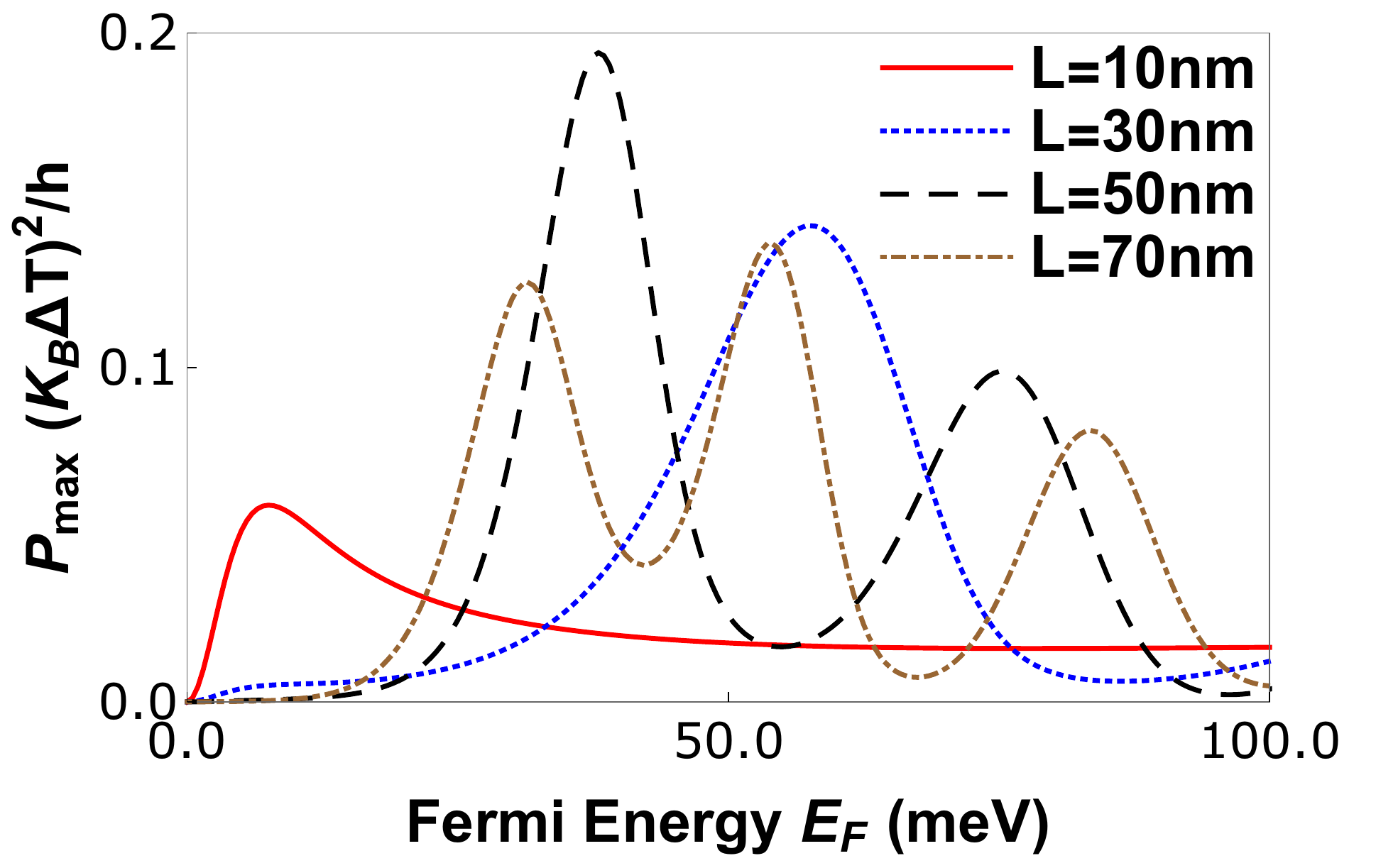}}
 \caption{Maximum Power ($P_{max}$) in units of ($(k_B\Delta T)^2/h$) at $30K$ for different lengths(L) of strained region with strain $= 50 meV$ and width(W) of strained region $=20 nm$.}
\end{figure}

  \begin{figure}
\centering {\includegraphics[width=0.4\textwidth]{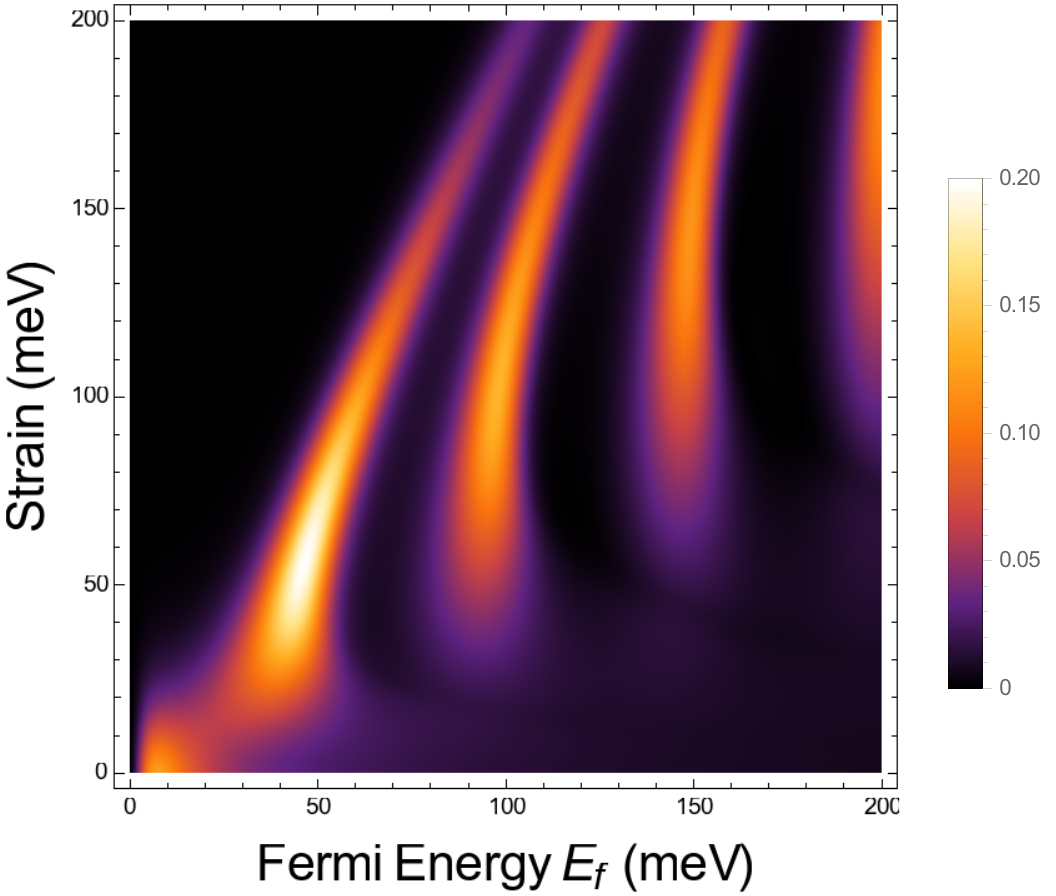}}
 \caption{Maximum Power ($P_{max}$) in units of ($(k_B\Delta T)^2/h$) at T= 30 K, where strain is along the y direction and Fermi energy $E_f$ is along the x direction with $L=40 nm$ and width $W=20 nm$.}
\end{figure}

At lower values of  strain ($s=50$ meV) our engine achieves maximum power, i.e., 0.2 $(k_b\Delta T)^2/h=0.057$ pico-Watts at $30 K$ for a $40 nm$ strained region, considering $\Delta T=1K$, see Figs.~4 and 5, which is more than two and three terminal quantum Hall heat engine at maximum power \cite{sothmann}.

The efficiency at maximum power $\eta (P_{max})$ is $0.1\eta_c$, which is also good enough as compared to the other QHE's, see Fig.~6. Efficiency at maximum power can also be increased to a large value(more than 0.4 $\eta_c$), as in Fig.~7, but then maximum power $P_{max}$ reduces to less than 0.03 $(k_b\Delta T)^2/h$. This is because while power depends on both Seebeck co-efficient and electrical conductance, see Eq.~(\ref{pmax}) the two factors so conspire to reduce the maximum power. On the other hand, the overall efficiency at maximum power again though dependent on  Seebeck co-efficient ($S$), conductance $G$ and thermal conductance $\kappa$, effectively increases with increasing strain. Individually,  S increases with increasing strain, while for $G$ and $\kappa$ it is the opposite.

 \begin{figure}
\centering {\includegraphics[width=0.44\textwidth]{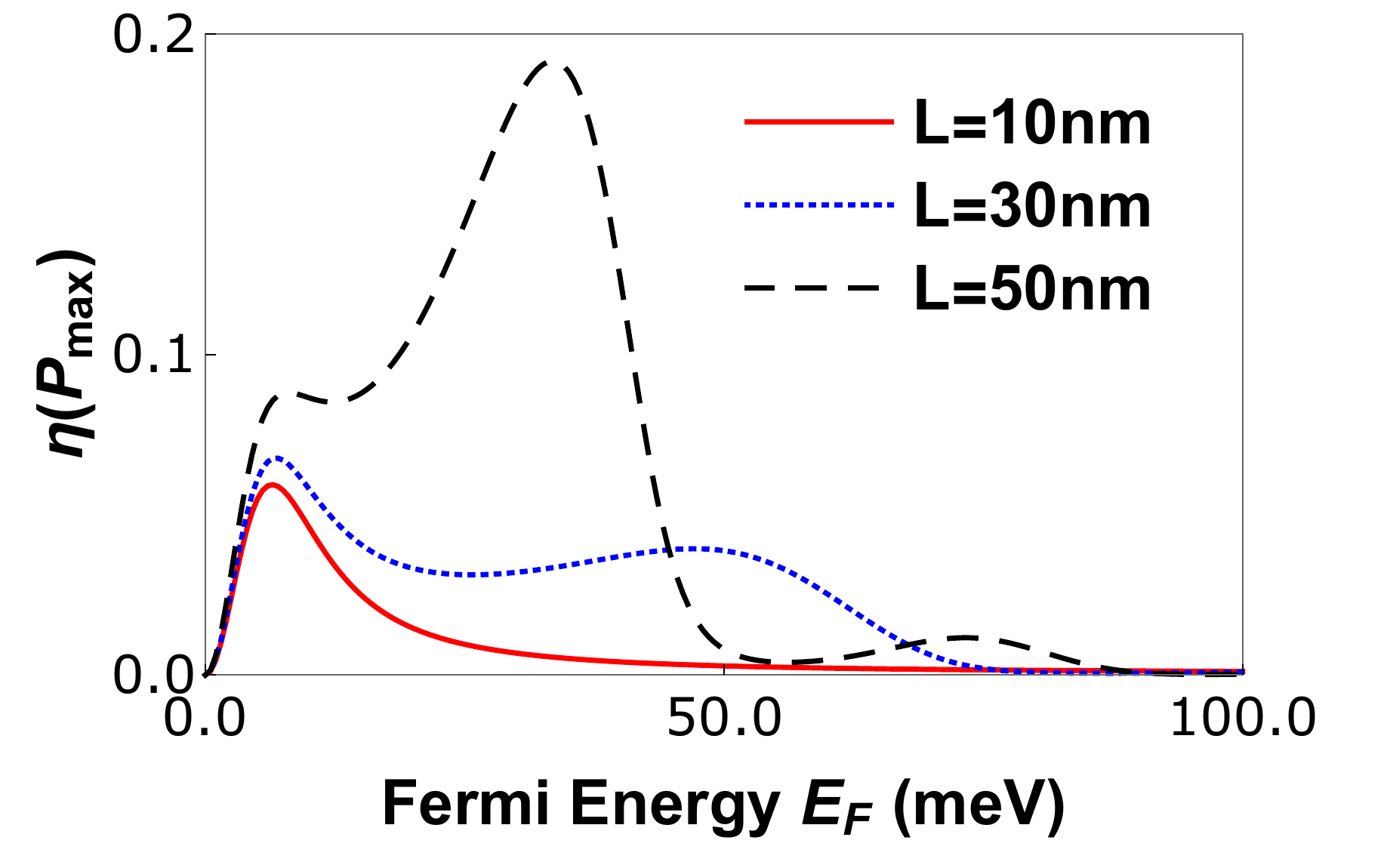}}
 \caption{Efficiency at maximum power in units of ($\eta_c$) at $30K$ with strain $= 50 meV$ and width $W=20 nm$.}
\end{figure}

  \begin{figure}
\centering{\includegraphics[width=0.4\textwidth]{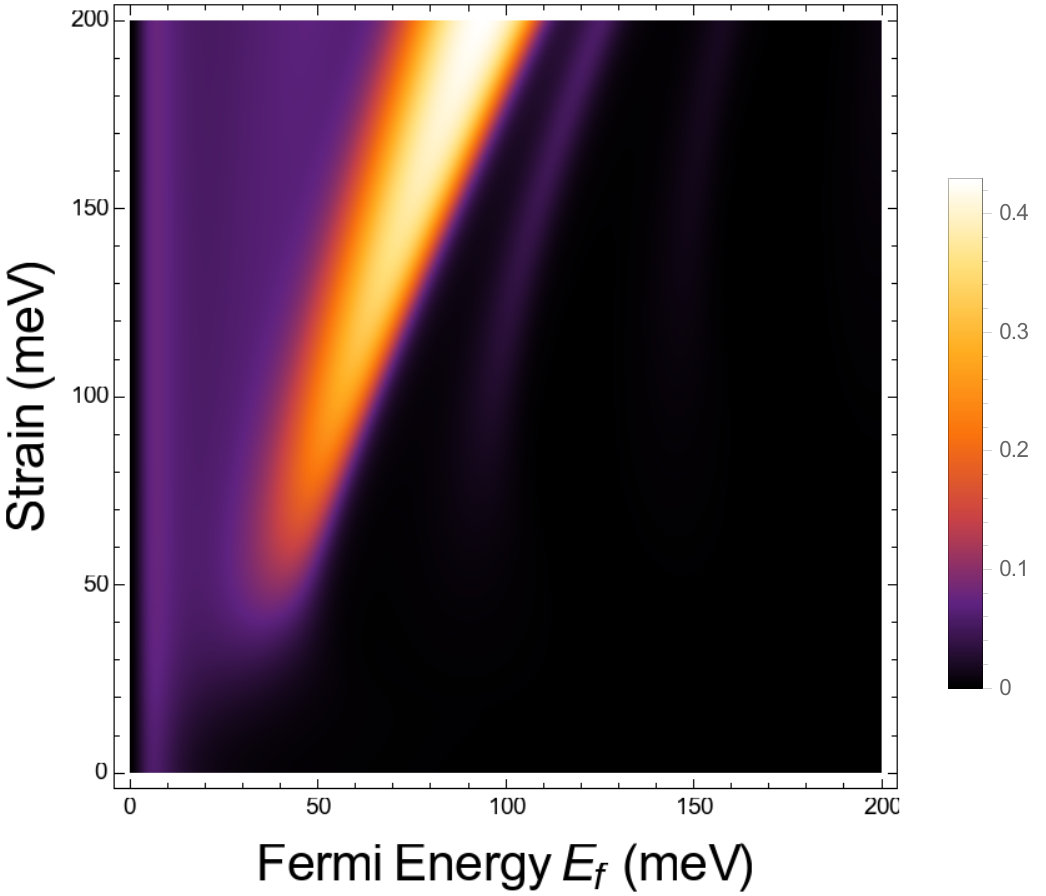}}
 \caption{Efficiency at maximum power ($\eta (P_{max})$) in units of ($\eta_c$) at $T=30K$ with width $W=20 nm$ and $L=40nm$. }
\end{figure}

\begin{figure}[b]
 \centering \subfigure[] {\includegraphics[width=0.48\textwidth]{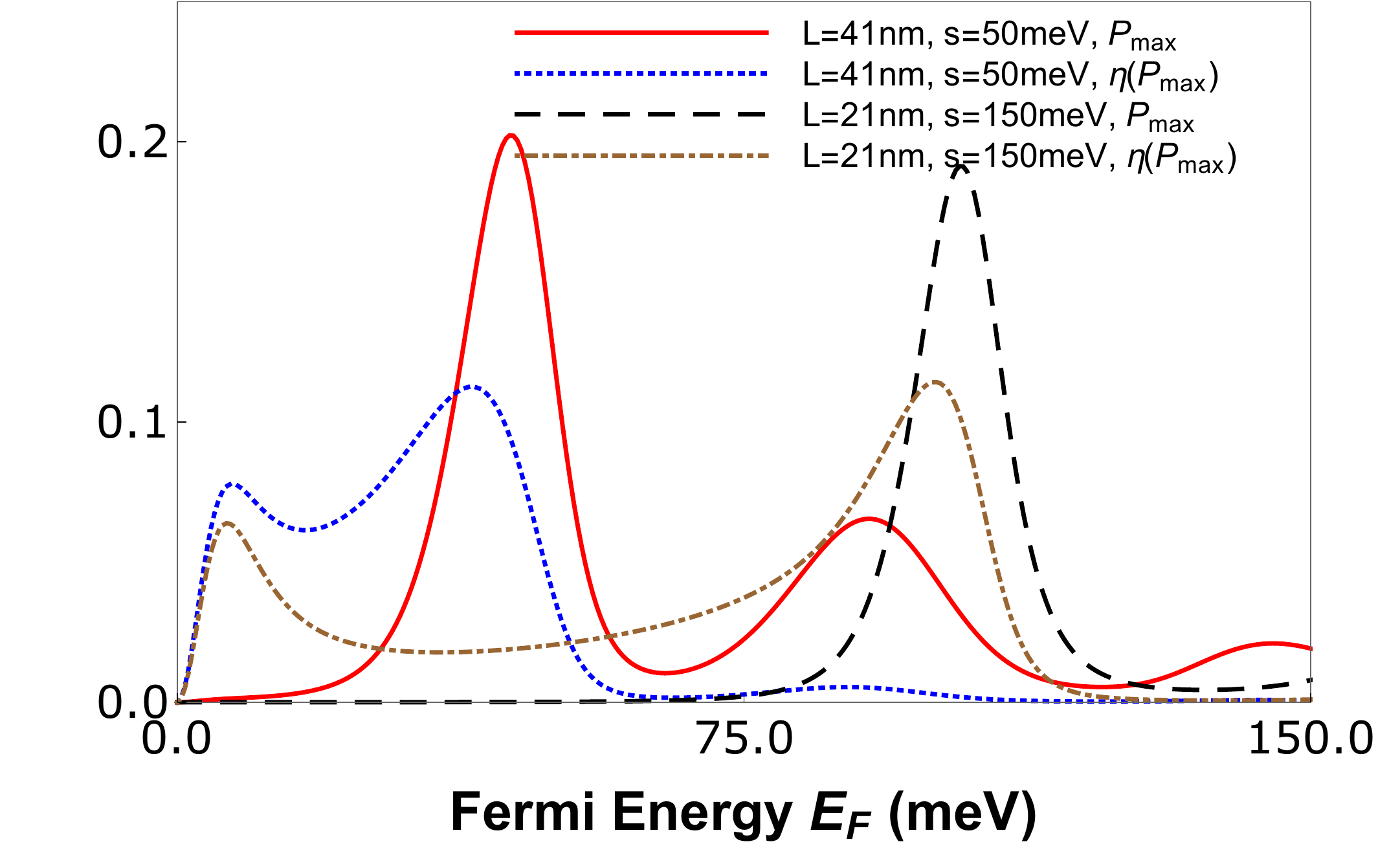}}
   \centering \subfigure[] {\includegraphics[width=0.48\textwidth]{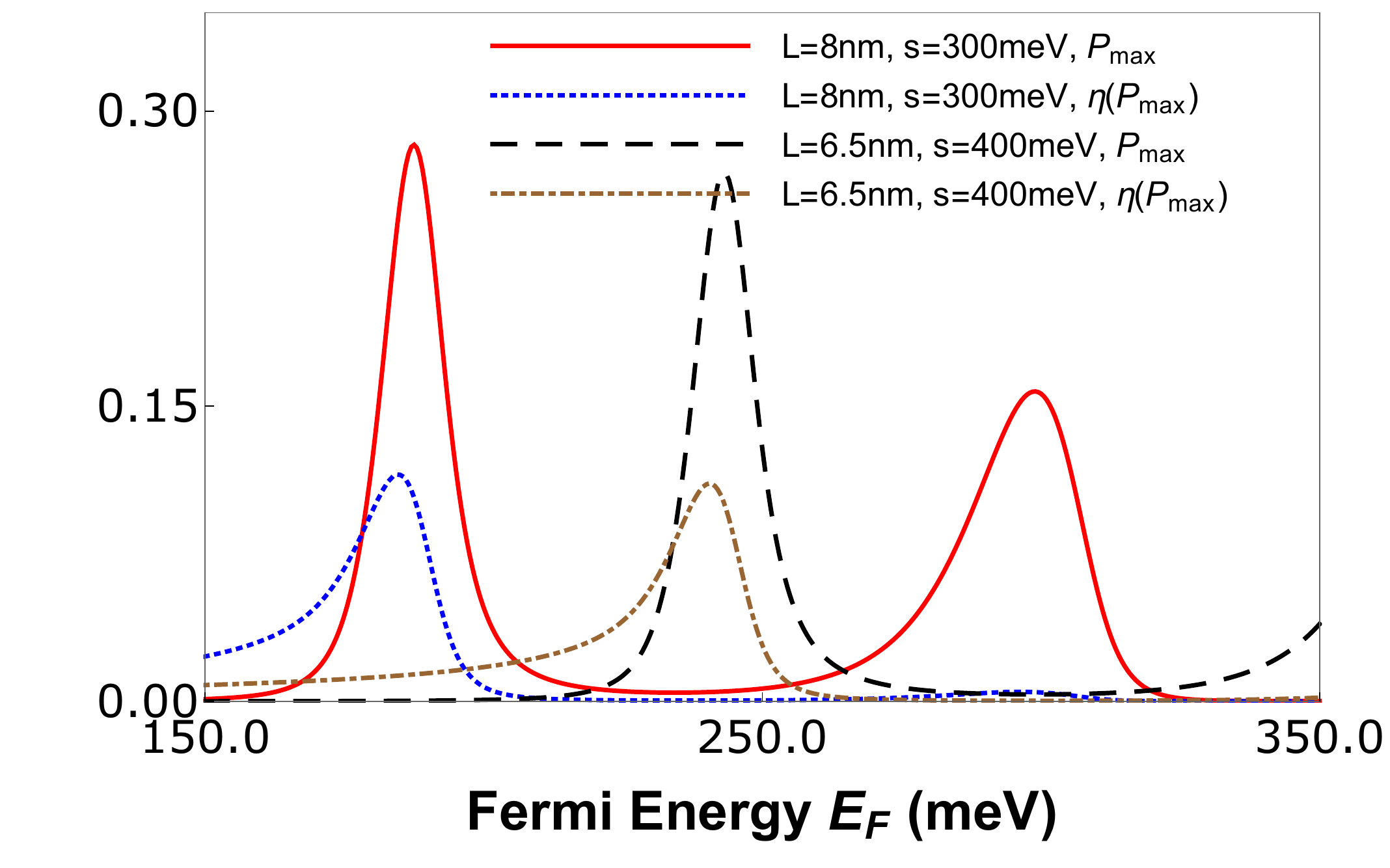}}
 \caption{(a) Maximum Power $P_{max}$ in units of ($(k_B\Delta T)^2/h$) and $\eta (P_{max})$ in units of ($\eta_c$) at $30 K$, for $v_f=10^6m/s$ and (b) Maximum Power $P_{max}$ in units of ($(k_B\Delta T)^2/h$) and $\eta (P_{max})$ in units of ($\eta_c$) at $30 K$, $v_f=6*10^5m/s$ with width $W=20 nm$.}
\end{figure}

Although the maximum efficiency $\eta (P_{max})$ and maximum power $P_{max}$ are good for this system, the dimension of the heat engine is large, equal to $20\times40$ $nm^2$. A effective QHE should deliver a high power with high efficiency and its dimensions should be as small as possible, so that in less area more number of nano heat engines can be fabricated, and thus good amount of power can be generated. From Fig.~8 (a) we see that with increasing strain ($150 meV$), while decreasing length ($L=21 nm$) large power and efficiency can be generated. The performance of the heat engine can be increased more by tuning one more variable, the Fermi velocity $v_f$. Till now, we have considered the Fermi velocity of Dirac electrons to be equal to $10^6$m/s, but increasing strain can reduce the Fermi velocity to $6*10^5$m/s \cite{behera}, then performance of the heat engine can be increased more, such as- maximum power as well as efficiency at maximum power both can be increased to a value as high as 0.268 $(k_b\Delta T)^2/h$ and 0.1$\eta_c$ respectively, see Fig.~8 (b). This can be understood better as, if 1$cm^2$ area is fabricated by this quantum nano heat engines in parallel, then $0.06 Watts$ total power can be generated with efficiency 0.1 $\eta_c$, which is better than quantum Hall heat engines but comparable to quantum dot heat engines, see Table I below.

Increasing temperature, Seebeck coefficient and electrical conductivity both can be increased to a large value with a maximum power more than 0.2$(k_b\Delta T)^2/h$ (at $v_f=10^6$ m/s) and efficiency at maximum power also more than $0.1\eta_c$. But then the phonon contribution to the thermal conductivity comes into play and that increases the thermal conductivity, implying a reduction in ZT, thermodynamic figure of merit. This reduces the efficiency at maximum power, though does not affect the power of the heat engine. We did not consider the phonon contribution, hence have restricted ourselves to an upper limit of  30K temperature at which the phonon contribution is neglected \cite{mazzamuto}.

\section{Co-efficient of Performance}
Finally we discuss the use of our model as a quantum refrigerator. As in our model external magnetic field is absent, so Time-Reversal (TR) symmetry  is not broken. The co-efficient of performance of the refrigerator is defined by the ratio of heat current extracted from the hot reservoir to the electrical power $\mathcal{P}$, such as -
\begin{eqnarray}
\eta^r=\frac{j^q}{\mathcal{P}}
\end{eqnarray} 
which is maximum, considering $j^q<0$ and $\mathcal{P}<0$, for -
\begin{eqnarray}
\mathcal{E}&=&\frac{L^{22}}{L^{21}}(-1-\sqrt{\frac{L^{11}L^{22}-L^{12}L^{21}}{L^{11}L^{22}}})\Delta T\\
\text{and,}\quad\eta^r_{max}&=&\eta^r_c\frac{\sqrt{ZT+1}-1}{\sqrt{ZT+1}+1},
\end{eqnarray} 
where $\eta^r_c=\frac{T}{\Delta T}$ is the efficiency of an ideal refrigerator. { For systems with broken TR symmetry, the upper bound of the refrigerator efficiency $\eta^r_{max}$ decreases from $\eta^r_c$ as the asymmetric parameter $x=TL^{12}/L^{21}$ deviates from 1 \cite{brandner}. For systems with conserved TR symmetry, the asymmetric parameter $x$ becomes unity, and the upper bound of the corresponding maximum efficiency  $\eta^r_{max}$ equals $\eta^r_c$. This is the advantage of systems with conserved TR symmetry, that it can work as both heat engine as well as a refrigerator with higher bound of efficiency, but for systems with broken TR symmetry, for refrigerator, this upper bound reduces from $\eta^r_c$}.

\begin{widetext}
{
\begin{center}
\begin{table}
\caption{How does the strained graphene QHE compare with related proposals?}
\begin{tabular}{ |c|p{3cm}|p{3cm}|p{3cm}|}
 \hline
Heat Engines&{Maximum Power $P_{max}$ in units of $(k_b\Delta T)^2/h$}& {Efficiency at maximum Power $\eta(P_{max})$}&{Power generated in 1 $cm^2$ area fabricated by nano engines}  \\ 
\hline
Quantum Hall Heat Engine(two terminal)\cite{sothmann}&0.14&0.10 $\eta_c$&0.04 Watts\\
\hline
Quantum Hall Heat Engine(three terminal) \cite{sothmann} &0.14&0.042 $\eta_c$&0.04 Watts\\
\hline
Chaotic Cavity\cite{buti}&0.0066&0.01 $\eta_c$&0.00189 Watts\\
\hline
Strained Graphene QHE&0.268&0.1 $\eta_c$& 0.06 Watts\\
\hline
\end{tabular}
\end{table}
\end{center}
}
\end{widetext}
\section{Conclusion}
We show here that strain acting solely can act as a QHE with better performance characteristics like high efficiency than most other QHE like quantum Hall heat engine, chaotic cavity QHE,  etc. It has some advantage over magnetically driven QHE. Application of magnetic field breaks the TR symmetry, which in turn reduces the performance of the system as a refrigerator. On the other hand, strain does not break TR symmetry, so our system can act as both heat engine as well as refrigerator \cite{brandner}.
In Table 1 we compare efficiency $\eta (P_{max})$, power $P_{max}$ and total power generated for some configured open QHE's. We see that our model system has excellent characteristics compared to other QHE's.
 This raises a question that perhaps large power and efficiency can also be found with different kind of strain patterns in multi-terminal graphene system, for which further investigations are needed.
    
 {\em  \underline {Acknowledgments:} }This work was supported by funds from Dept. of Science and Technology (SERB), Govt. of India, Grant No. EMR/2015/001836.


\begin{thebibliography}{99}
\bibitem{selman} K.A. Muttalib and Selman Hershfield, Nonlinear Thermoelectricity in Disordered Nanowires, Phys. Rev. Applied 3, 054003 (2015).
\bibitem{benenti} G. Benenti, G. Casati, K. Saito, and R. S. Whitney, Fundamental aspects of steady state conversion of heat to work at the nanoscale, arXiv: 1608.05595v3.
\bibitem{ron} Ronnie Kosloff and Amikam Levy, Quantum Heat Engines and Refrigerators: Continuous Devices, Annu. Rev. Phys. Chem. 2014. 65:365-93.
\bibitem{prosen} T. Prosen, Revisiting Thermodynamic Efficiency, Physics 6, 16 (2013).
\bibitem{munoz} F. J. Pena and E. Munoz, Magnetostrain-driven quantum engine on a graphene flake, Phys. Rev. E 91, 052152 (2015).
\bibitem{yun} Jeong Yun Kim and Jeffrey C. Grossman, High-Efficiency Thermoelectrics with Functionalized Graphene, NanoLett. 2015, 15, 2830.
\bibitem{anno} Y. Anno, et. al., Enhancement of graphene thermoelectric performance through defect engineering, 2D Mater. 4 (2017) 025019.
\bibitem{tran} V.-T. Tran, et. al., Optimizing the thermoelectric performance of graphene nanoribbons without degrading the  electronic properties, Scientific Reports 7, 2313 (2017).
\bibitem{sadeghi} H. Sadeghi, et. al., Enhancing the thermoelectric figure of merit in engineered graphene nanoribbons, Beilstein J. Nanotechnol. 2015, 6, 1176.
\bibitem{dragoman} D. Dragoman and M. Dragoman, Giant thermoelectric effect in graphene, APPLIED PHYSICS LETTERS 91, 203116 (2007).
\bibitem{gahari} F. Ghahari, et. al., Enhanced Thermoelectric Power in Graphene: Violation of the Mott Relation by Inelastic Scattering, Phys. Rev. Lett. 116, 136802 (2016).
\bibitem{dolphas} M. C. Nguyen, et. al., Enhanced Seebeck effect in graphene devices by strain and doping engineering, Physica E. 73, 207 (2015).
\bibitem{ramshetti} Babak Zare Rameshti and Ali G. Moghaddam, Spin-dependent Seebeck effect and spin caloritronics in magnetic graphene, Phys. Rev. B 91, 155407 (2015).
\bibitem{benetti} G. Benenti, K. Saito, G. Casati, Thermodynamics Bounds on Efficiency for systems with Broken Time-Reversal Symmetry, PRL 106, 230602 (2011).
\bibitem{castro} V. M. Pereira and A. H. Castro Neto, Strain Engineering of Graphene's Electronic Structure, PRL 103, 046801 (2009).
\bibitem{buti} B. Sothmann, R. Sanchez, A. N. Jordan, and M. Buttiker, Phys. Rev. B 85, 205301 (2012).
\bibitem{sothmann} Patrick P. Hofer and Bjorn Sothmann, Quantum heat engines based on electronic Mach-Zehnder interferometers, Phys. Rev. B 91, 195406 (2015).
\bibitem{jordan} A. N. Jordan, B. Sothmann, R. Sanchez, and M. Buttiker, Phys. Rev. B 87, 075312 (2013).
\bibitem{pena} E. Munoz, F. J. Pena, Magneticlly driven quantum heat engine, Phys. Rev. E 89, 052107 (2014).
\bibitem{behera} Harihar Behera and Gautam Mukhopadhyay, Fermi Velocity Modulation in Graphene by Strain Engineering,  AIP Conference Proceedings, 1512, 1 (2013).  
\bibitem{mazzamuto} F. Mazzamuto, et. al., Enhanced thermoelectric properties in graphene nanoribbons by resonant tunneling of electrons, Phys. Rev. B 83, 235426 (2011).
\bibitem{brandner} Kay Brandner, et. al., Strong Bounds on Onsager Coefficients and Efficiency for Three-Terminal Thermoelectric Transport in a Magnetic Field, Phys. Rev. Lett. 110, 070603 (2013).
\bibitem{perera} V. M. Pereira and A. H. Castro Neto, Tight-binding approach to uniaxial strain in graphene, Phys. Rev. B. 80, 045401 (2009).
 \end{thebibliography}
\end{document}